\documentclass{article}
\usepackage{graphicx}
\title{Quantum Machines}
\author{Mark Hillery  \\ Department of Physics \\ Hunter College of CUNY \\ 695 Park Avenue \\
New York, NY 10061, USA
\and
Vladimir Bu\v{z}ek \\Research Center for Quantum Information \\
Slovak Academy of Sciences \\Dubravska cesta 9 \\ 845 11 Bratislava, Solvakia}
\begin{document}
\maketitle
\begin{abstract}
We discuss quantum information processing machines.  We start with single purpose machines
that either redistribute quantum information or identify quantum states.  We then move on to
machines that can perform a number of functions, with the function they perform being determined
by a program, which is itself a quantum state.  Examples of both deterministic and probabilistic 
programmable machines are given, and we conclude with a discussion of the utility of 
quantum programs.
\end{abstract}

\section{Introduction}
Quantum information is information stored in a quantum mechanical system \cite{nielsenbook}.  
The systems themselves
are either collections of qubits, two-dimensional systems, or qudits, $D$-dimensional systems.  For 
example, a qubit exists in a two dimensional space spanned by two orthonormal states, which are
denoted by $|0\rangle$ and $|1\rangle$.  While a classical bit can be either $0$ or $1$, a qubit can
be in any superposition of $|0\rangle$ and $|1\rangle$, i.e. $|\psi\rangle = \alpha |0\rangle + \beta
|1\rangle$, where $\alpha$ and $\beta$ are complex numbers satisfying $|\alpha |^{2} + |\beta |^{2}
=1$.  This fact leads to large differences in the properties of classical and quantum information.  Once
one has information is in the form of qubits of qudits, one would like to do something with it.  Ultimately,
in order to extract the information stored in the system, it will be necessary to measure the it, but
before doing so, it is usually useful to perform some operations on the system.  For example, one
might want to (approximately)
copy it, compare it with known quantum states or with information stored in other quantum systems,
or perform an operation on it.  The devices that accomplish these tasks are quantum machines.
They can either be single-purpose or programmable and able to perform many tasks.

In the case of programmable systems, the program can be either quantum or classical.  Examples
of a classical programs are a sequence of laser pulses to put a molecule into a particular quantum
state or a sequence of radio frequency pulses to control the dynamics of a system of nuclear
spins \cite{dalessandro}.  Here, we want to consider programs that are themselves states of
quantum systems, i.e. we want to perform quantum programming.  Quantum programs have some
properties that classical ones do not.  For example, they can exist in superposition states, which 
means that a quantum processor can perform several programs simultaneously.  In addition, 
quantum programs are necessary when the information on which the program is based is, in fact,
quantum information.  We shall provide examples of each of this situations. 

As we shall see, in a number of cases, it is not possible to perform the desired task perfectly.  For
example, the no-cloning theorem prohibits the perfect copying of quantum information \cite{wooters}. 
We are then faced with deciding how to perform the task as best as we can.  This usually amounts
to adopting one of two possible strategies.  The first is to produce an output state that is as close
as possible to the ideal output, that is we approximately perform the desired task.  The second is
to perform the task with some probability in such a way that we know 
whether the proper task has been performed or not.  In that case, the task has been
performed probabilistically.   

We shall begin by discussing single-purpose machines, in particular cloners and machines that
perform the discrimination of quantum states.  Next, we shall move on to a general discussion
of programmable machines and outline the deterministic and probabilistic 
approaches to them.  We will present
a no-go theorem that shows that a deterministic programmable machine cannot be universal.  We then
consider the problem of implementing a one-parameter unitary group on a programmable machine
and discuss the approximate and probabilistic strategies.  We also show how, in the probabilistic
case, to increase the probability of success by increasing the size of the program space.  Finally,
we make a case for why quantum programs are useful.

\section{Singe-purpose machines}
\subsection{Cloners}
As was mentioned in the Introduction, the no-cloning theorem prohibits the perfect copying of
quantum information.  The theorem states that if we have a quantum system in the state $|\psi\rangle$,
we cannot build a machine whose output will be $|\psi\rangle |\psi\rangle$, for all $|\psi\rangle$.
The proof relies simply on the linearity of quantum mechanics \cite{wooters}.  If the machine 
clones each vector of a basis for the input space, then its action is completely determined, and it is
incompatible with the transformation $|\psi\rangle \rightarrow |\psi\rangle |\psi\rangle$.

An approximate cloner for qubits can be constructed from four Controlled-NOT gates and three
qubits (see Figure 1) \cite{buzek1,braunstein,gisin}.  A Controlled-NOT gate is a two-qubit gate.
The first qubit is the control qubit, and the second is the target qubit.  If the state of the control 
qubit is $|0\rangle$, then nothing happens to the states of either the control or the target qubits.
However, if the state of the control qubit is $|1\rangle$, then the state of the control qubit is again
unchanged, but the state of the target qubit is flipped, i.e. if it was $|0\rangle$ it becomes $|1\rangle$,
and vice versa.  Returning now to the cloner, we have that the first qubit, in the state 
$|\psi\rangle = \alpha |0\rangle + \beta |1\rangle$,
where $\{ |0\rangle , |1\rangle \}$ is the orthonormal qubit basis, is the one to be copied, and 
second qubit will become the copy.  In order to see how this works, define the two two-qubit states
\begin{eqnarray}
|\Xi_{00}\rangle & = &  \frac{1}{\sqrt{2}} (|0\rangle |0\rangle + |1\rangle |1\rangle )  \nonumber  \\
|\Xi_{0x}\rangle & = & \frac{1}{\sqrt{2}} |0\rangle (|0\rangle + |1\rangle ) .
\end{eqnarray}
We now note that if qubit $1$ is in the state $|\psi\rangle_{1}$ and qubits $2$ and $3$ are in one of
the two states above, then the cloning circuit will implement the following transformations
\begin{eqnarray}
|\psi\rangle_{1}|\Xi_{00}\rangle_{23} & \rightarrow & |\psi\rangle_{1}|\Xi_{00}\rangle_{23}  \nonumber \\
|\psi\rangle_{1}|\Xi_{0x}\rangle_{23} & \rightarrow & |\psi\rangle_{2}|\Xi_{00}\rangle_{13}  .
\end{eqnarray}
Examining these equations, we see that in the first the quantum information from the first qubit 
appears in output $1$, and in the second it appears in output $2$.  This suggests that if instead
of sending either $|\Xi_{00}\rangle$ or $|\Xi_{0x}\rangle$ into inputs $2$ and $3$, we send in a
linear combination of them, some of the quantum information from qubit $1$ will appear in output
$1$ and some of it will appear in output $2$, thereby cloning the state.  This is, in fact, exactly what
happens.  If we choose
\begin{equation}
|\Psi\rangle_{23} = c_{0}|\Xi_{00}\rangle_{23}+c_{1}|\Xi_{0x}\rangle_{23} ,
\end{equation}
as the input state for qubits $2$ and $3$, with $c_{0}$ and $c_{1}$ real for simplicity,
then the reduced density matrices for outputs $1$ and $2$ are
\begin{eqnarray}
\rho_{1}^{(out)} & = & (c_{1}^{2} + c_{0}c_{1}) |\psi\rangle \langle\psi | + \frac{c_{1}^{2}}{2}I \nonumber \\
\rho_{2}^{(out)} & = & (c_{0}^{2} + c_{0}c_{1}) |\psi\rangle \langle\psi | + \frac{c_{0}^{2}}{2}I  .
\end{eqnarray}
Note that by choosing $c_{0}$ and $c_{1}$ we can control how much information about $|\psi\rangle$
goes to which output.  In particular, if we choose $c_{0}=c_{1}=1/\sqrt{3}$, then the information is
divided equally, and we find that
\begin{equation}
\rho_{1}^{(out)}=\rho_{2}^{(out)}=\frac{5}{6}|\psi\rangle \langle\psi | + \frac{1}{6} |\psi_{\perp}\rangle
\langle \psi_{\perp}| ,
\end{equation}
where $|\psi_{\perp}\rangle$ is the qubit state orthogonal to $|\psi\rangle$.  Therefore, the fidelity
of the cloner output $\rho_{1}^{(out)}$ (or $\rho_{2}^{(out)}$, since they are the same in this case) 
to the ideal output, $|\psi\rangle$, which is given by $\langle\psi |\rho^{(out)}_{1}|\psi\rangle$,
is $5/6$.  A fidelity of one would imply perfect cloning, so what we have here is a device that 
produces two copies of the input qubit that are pretty good approximations to it. 

\begin{figure}
\includegraphics{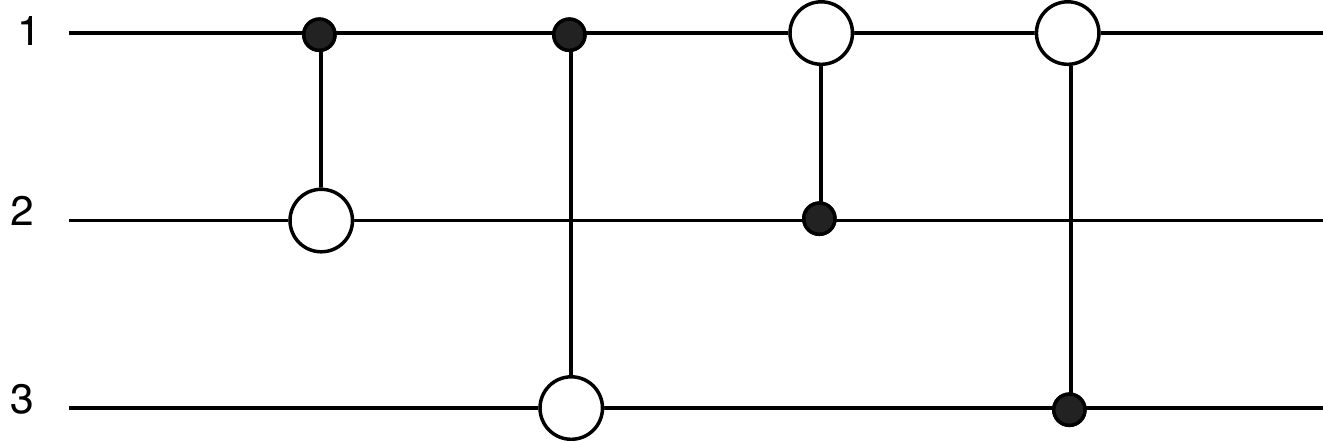}
\caption{A circuit for an approximate quantum cloner with three qubits and four Controlled-NOT
gates.  The qubit to be copied goes into input 1, and the copies come out in outputs 1 and 2.
The anti-clone comes out of output 3.  In the symbol for the Controlled-NOT gate, the filled 
circle indicate the control qubit, and the open circles indicate the target qubit.}
\end{figure}

Note that the cloner employs three qubits, and we have only discussed the final state of two of
them.  One might wonder if the output state of the third qubit is of interest.  The answer ``yes.''
Its state is the best approximation to the state orthogonal to that of the input qubit that can be
realized.  A machine that sends a qubit in an arbitrary input state $|\psi\rangle$ into the
state orthogonal to it, $|\psi_{\perp}\rangle$, is known as a universal-NOT, or UNOT, gate, and
this transformation is also impossible to perform exactly \cite{werner1,werner2}.  It can, however,
be performed approximately, and the best fidelity that can be obtained is $2/3$.  This can be
achieved by measuring the original qubit along an arbitrary axis and then producing an output qubit 
whose state is orthogonal to the state obtained as the result of the measurement.  For example, if
one measured along the $z$ axis and found the result $+z$, one would create a qubit in the $-z$
direction.  The same result can be achieved by taking the third qubit, the one which is not a clone,
from the output of a quantum cloning machine.  This output qubit is sometimes referred to as an
anticlone.  

There have been a number of realizations of a quantum cloning machine, most based on a device
known as an optical parametric amplifier \cite{demartini1, demartini2,lamas}.  This device takes
one photon at frequency $2\omega$ and converts it into two photons at frequency $\omega$.  A
strong beam at $2\omega$ will  amplify a weaker beam at frequency $\omega$ via
stimulated emission.  When this device is used as a cloner, the qubits are the polarization states of
the photons.  A photon in an arbitrary polarization state at frequency $\omega$ will produce
three photons, all at frequency $\omega$, at the output.  Two of them will approximate clones, 
and the third will be an approximate anticlone.

A second cloning strategy is a probabilistic one \cite{duan}.  In this case, one wants to clone
a quantum state that is selected from a known, finite set of states.  For simplicity, let us assume
that this set contains two elements, $|\psi_{1}\rangle$ and $|\psi_{2}\rangle$.  Our machine is
then to do the following.  Given an input qubit that is in either the state $|\psi_{1}\rangle$ or 
$|\psi_{2}\rangle$, we don't know which, it is to produce two copies of the input state.  If the two
input states are not orthogonal, this cannot be done perfectly.  It can, however, be done 
probabilistically.  The machine either produces two perfect copies of the input, or it fails, and it
tells us which of these two possibilities has occurred.  The probability of successfully cloning
the input is given by
\begin{equation}
p_{succ} = \frac{1}{1+ |\langle \psi_{1}|\psi_{2}\rangle |} .
\end{equation}
Note that this is one if the states are orthogonal, and decreases as their overlap increases.

\subsection{State discriminators}
The problem in quantum state discrimination is, given a particle in an unknown state selected from
a known set of states, determine the quantum state of the particle \cite{bergou,chefles}.  
If the set of possible states
contains states that are not orthogonal to each other, then this cannot be done
perfectly.  Again, for simplicity, let us assume that our set of possible states contains only two
states, $|\psi_{1}\rangle$ and $|\psi_{2}\rangle$.  

We shall again explore two strategies for accomplishing this task.  The first is the minimum-error
strategy, which is a strategy that approximately discriminates the two states \cite{helstrom}.  
It can make mistakes, but the probability of making a mistake is minimized.  A machine
that implements this strategy is given an input, which is equally likely to be $|\psi_{1}\rangle$ or
$|\psi_{2}\rangle$, and it then tells us which of the two states it was given.  The probability of 
the output being incorrect is
\begin{equation}
p_{err} = \frac{1}{2}(1- \sqrt{1-|\langle\psi_{1}|\psi_{2}\rangle |^{2}}) . 
\end{equation}
Note that when the states are orthogonal, this is zero, but that it increases as the overlap between
the states increases.  The second strategy is known as unambiguous state discrimination
\cite{ivanovic,dieks,peres}.  In this case our machine has three outputs, one corresponding to
state $1$, one corresponding to state $2$, and a third corresponding to failure.  This machine
will never incorrectly identify a state, but it may fail.  For example, if the input is in $|\psi_{1}\rangle$,
the machine will either tell us the input was in state $|\psi_{1}\rangle$, or fail, but it will never 
tell us the input was in state $|\psi_{2}\rangle$.  Assuming that each input state is equally likely, the
probability of successfully identifying the state is
\begin{equation}
p_{succ}=1-|\langle\psi_{1}|\psi_{2}\rangle | .
\end{equation}
As has been the case before, the probability of successful identifying the input state is one if the
states are orthogonal, and decreases as the overlap of the states increases.

\section{Programmable machines}
We now want to consider programmable quantum machines, which we shall often refer to as
quantum processors \cite{buzek2}.  These have two inputs, one for the 
data, which is to be acted upon, and one for the program, which will specify the operation to be
performed on the data.  Both the data and the program are quantum states.  In particular, the
processor is a unitary operator acting on the Hilbert space $\mathcal{H}_{d}\otimes\mathcal{H}_{p}$,
where $\mathcal{H}_{d}$ is the data Hilbert space and $\mathcal{H}_{p}$ is the program 
Hilbert space.  The machine can
act in either a deterministic or probabilistic fashion.  In the case of a deterministic machine, we 
always accept the output, and the action of the machine on the data state is described by a 
trace-preserving completely positive map, which is a result of tracing out the program
state output.  In the case of a probabilistic machine, we measure the program state output, and only
accept the data state output if a particular result is obtained.  We shall examine both scenarios.

It is, perhaps, best to begin with an example \cite{hillery1}.  
Let us go back and consider the three-qubit circuit
for the approximate cloner.  Qubit $1$ will now be our data state, and qubits $2$ and $3$ will be 
our program.  We will denote the data state by $|\psi\rangle_{1}$ and the program state by
$|\Xi\rangle_{23}$.  Define the two-qubit Bell states to be
\begin{eqnarray}
|\Psi_{\pm}\rangle & = & \frac{1}{\sqrt{2}}(|00\rangle \pm |11\rangle ) \nonumber  \\
|\Phi_{\pm}\rangle & = & \frac{1}{\sqrt{2}}(|01\rangle \pm |10\rangle ) .
\end{eqnarray}
If these states are used as programs in our processor, we find that
\begin{eqnarray}
|\psi\rangle_{1}|\Psi_{+}\rangle_{23} & \rightarrow & |\psi\rangle_{1}|\Psi_{+}\rangle_{23} \nonumber \\
|\psi\rangle_{1}|\Psi_{-}\rangle_{23} & \rightarrow & \sigma_{z}|\psi\rangle_{1}|\Psi_{-}\rangle_{23}
\nonumber \\
|\psi\rangle_{1}|\Phi_{+}\rangle_{23} & \rightarrow & \sigma_{x}|\psi\rangle_{1}|\Phi_{+}\rangle_{23}
\nonumber \\
|\psi\rangle_{1}|\Psi_{-}\rangle_{23} & \rightarrow & (-i\sigma_{y})|\psi\rangle_{1}|\Psi_{-}\rangle_{23} ,
\end{eqnarray}
where $\sigma_{x}$, $\sigma_{y}$, and $\sigma_{z}$ are the Pauli matrices.  If we choose the
program
\begin{eqnarray}
|\Xi\rangle_{23} & = & c_{0}|\Psi_{+}\rangle_{23} + c_{1} |\Phi_{+}\rangle_{23} \nonumber \\
& & + c_{2} |\Phi_{-}\rangle_{23} + c_{3} |\Psi_{-}\rangle_{23} ,
\end{eqnarray}
then operating our machine in the deterministic mode, by tracing out the program state output, 
we obtain for the data state output
\begin{eqnarray}
\label{procmap}
\rho_{1}^{(out)} & = & |c_{0}|^{2}\rho_{1}^{(in)} + |c_{1}|^{2} \sigma_{x}\rho_{1}^{(in)}\sigma_{x}
\nonumber \\
& & |c_{2}|^{2} \sigma_{y}\rho_{1}^{(in)}\sigma_{y} + |c_{3}|^{2} \sigma_{z}\rho_{1}^{(in)}\sigma_{z} .
\end{eqnarray}
In the above equation, we have set $\rho_{1}^{(in)}=|\psi\rangle_{1}\langle\psi |$.  Examining 
the output state, we see that this circuit can implement a number of quantum channels: the
bit-flip channel ($c_{2}=c_{3}=0$), which flips a bit, with a certain probability, the phase-flip
channel ($c_{1}=c_{2}=0$), which sends $|0\rangle \rightarrow |0\rangle$ and $|1\rangle
\rightarrow -|1\rangle$, with a certain probability, and the depolarizing channel 
($c_{1}=c_{2}=c_{3}$), in which the input state is replaced by the completely mixed state, with a
certain probability \cite{nielsenbook}.   The same processor can be used in the probabilistic mode.
Suppose we want to implement the operator $A=I-2|\phi\rangle\langle\phi |$ on the data state,
where $|\phi\rangle$ is a specified one-qubit state.  The operator $A$ is similar to $\sigma_{z}$,
but instead of flipping the phase of the state $|1\rangle$, it flips the phase of the state $|\phi\rangle$.
Defining the two-qubit operator, $U_{0}$,
\begin{eqnarray}
U_{0}|00\rangle = - |10\rangle & U_{0}|10\rangle = -|11\rangle  \nonumber  \\
U_{0}|01\rangle = |00\rangle & U_{0}|11\rangle = |01\rangle  ,
\end{eqnarray}
we choose for our program state 
\begin{equation}
|\Xi\rangle_{23}= \frac{1}{\sqrt{2}}U_{0}(|\phi\rangle_{2}|\phi_{\perp}\rangle_{3} 
+ |\phi_{\perp}\rangle_{2} |\phi\rangle_{3} ) ,
\end{equation}
where $|\phi_{\perp}\rangle$ is the qubit state orthogonal to $|\phi\rangle$.  At the program state
output, we project onto the state $(|\Phi_{+}\rangle_{23} + |\Phi_{-}\rangle_{23} 
+ |\Psi_{-}\rangle_{23})/\sqrt{3}$, and if we get one, we keep the data state output.  This will 
happen with a probability of $1/3$, independent of the state $|\phi\rangle$.  If we do get one, 
then the data state output will be in the state $A|\psi\rangle$.

Now, let us return to deterministic processors and examine the resources that are necessary in
order to implement a given set of operations on the data.  Suppose that our data state is a qubit,
and we want to implement a one-parameter unitary group $U(\alpha )=\exp (i\alpha \sigma_{z})$, 
where $0\leq \alpha < 2\pi$, on it.  We want to encode the angle $\alpha$ in the program state.  It
turns out that this cannot be done with a finite dimensional program space, due to a no-go
theorem due to Nielsen and Chuang \cite{nielsen}.  It states that if the program $|\Xi_{1}\rangle$
implements the unitary operator $U_{1}$ on the data state, and $|\Xi_{2}\rangle$ implements the 
unitary operator $U_{2}$, then $\langle\Xi_{1}|\Xi_{2}\rangle = 0$.  This implies that for every unitary
operator that the processor can implement on the data state, we need an extra dimension in the 
program space.  In the case of our one-paramenter group, there are an infinite number of operators,
so it clearly cannot be implemented on a processor with a finite-dimensional program space.

Given this result, we can adopt the same strategies we did in the case of single-purpose machines
that were prohibited by a no-go theorem.  We have already seen in our example, that a probabilistic
machine has no problem implementing an infinite number of operations.  The only remaining issue
in that case is figuring out how to make the success probability as large as possible.  This we shall
address shortly.  The other strategy is to construct a machine that carries out a set of operations
approximately.  It is this type of machine we shall discuss now \cite{hillery2}.

We have seen that deterministic processors implement trace-preserving, completely positive maps.
Therefore, when considering approximate deterministic processor, we need to have some kind of
a measure of how close two such maps are.  We shall use the process fidelity \cite{raginsky},
which has a number of useful properties \cite{gilchrist}.  Let $T_{1}$ and $T_{2}$ be two
trace-preserving, completely positive maps, mapping the space, $\mathcal{B}(\mathcal{H})$, of linear
operators on a $D$-dimensional Hilbert space, $\mathcal{H}$, into itself.  The Jamiolkowski
isomorphism associates a density matrix on $\mathcal{H}\otimes\mathcal{H}$ to each 
trace-preserving, completely positive map on $\mathcal{B}(\mathcal{H})$.  Letting $\{ |j\rangle |
j=1,2,\ldots D\}$ be an orthonormal basis for $\mathcal{H}$, define the maximally entangled state,
$|\Phi\rangle$ in $\mathcal{H}\otimes\mathcal{H}$
\begin{equation}
|\Phi\rangle = \frac{1}{\sqrt{D}} \sum_{j=1}^{D} |j\rangle |j\rangle .
\end{equation}
The density matrix associated with the trace-preserving, completely positive map, $T$, is
\begin{equation}
\rho = (\mathcal{I}\otimes T)(|\Phi\rangle\langle\Phi |) ,
\end{equation}
where $\mathcal{I}$ is the identity map.  If $\rho_{1}$ is the density matrix associated with $T_{1}$
and $\rho_{2}$ is the density matrix associated with $T_{2}$, the the process fidelity between
$T_{1}$ and $T_{2}$ is
\begin{equation}
F_{proc}(T_{1},T_{2})= [ {\rm Tr}(\sqrt{\rho_{1}}\rho_{2}\sqrt{\rho_{1}})^{1/2}]^{2}  .
\end{equation}

We will not discuss the case of a general approximate processor, but will look at a specific type.
Suppose we have a processor that is a controlled-U gate.  That means that if our program
space, $\mathcal{H}_{p}$,  has dimension $N$, there is an orthonormal basis of $\mathcal{H}_{p}$,
$\{ |j\rangle |j=1,2,\ldots N \}$ such that the processor acts as follows
\begin{equation}
|\psi\rangle |j\rangle \rightarrow U_{j}|\psi\rangle |j\rangle ,
\end{equation}
where $U_{j}$ is a unitary operator on the data space.  Therefore, this processor implements the
set of unitary operators $S_{u}=\{ U_{j} | j=1,2,\ldots N \}$ perfectly.  Now suppose we want to use this
processor to approximate another unitary operator $U$, which is not in $S_{u}$.  We want to 
choose a program state that maximizes the process fidelity between the map it generates and $U$.
What one finds is that the best program is one of the basis states $|j\rangle$, and it is the one
for which $|{\rm Tr}(U^{\dagger}U_{j})|$ is a maximum.  In this case, one simply chooses the
unitary operator in $S_{u}$ that is closest to $U$ and implements that operator.  Using a program
state that is a superposition of different basis vectors does not help. 

Let us look at an example of this situation.  We wish to implement the operator, for $0\leq \theta <2\pi$, 
\begin{equation}
U(\theta ) = \exp \left[ \frac{i\pi}{2} (e^{-i\theta }\sigma^{+} + e^{i\theta }\sigma^{-})\right]  ,
\end{equation}
on our data state, which is a qubit.  Here $0\leq \theta <2\pi$, and $\sigma^{\pm} 
= (\sigma_{x}\pm i\sigma_{y})/2$.  Our
program state has dimension $N$, and an orthonormal basis $\{ |j\rangle | j=1,2\ldots N\}$.
Define the operators $E_{\pm}$ on the program space by $E_{\pm}|j\rangle = |j\pm 1\rangle$,
where the addition and subtraction are modulo $N$.  Now let the overall processor unitary
operator, which acts on the tensor product of the data and program spaces, be
\begin{equation}
G=\exp  \left[ \frac{i\pi}{2}( \sigma^{+}E_{-}+\sigma^{-}E_{+})\right] ,
\end{equation}
and consider the program states
\begin{equation}
|\theta\rangle = \frac{1}{\sqrt{N}}\sum_{j=0}^{N-1}e^{-ij\theta}|j\rangle .
\end{equation}
When $\theta = \theta_{m}=2\pi m/N$, for $m$ an integer between $0$ and $N-1$, we find that
\begin{equation}
G(|\psi\rangle |\theta_{m}\rangle ) = U(\theta_{m})|\psi\rangle |\theta_{m}\rangle ,
\end{equation}
where $|\psi\rangle$ is a general qubit state.  Therefore, this processor implements the operations
$U(\theta_{m})$ perfectly.  Now suppose we want to implement $U(\theta )$ for a value of $\theta$
that is not one of the $\theta_{m}$.  The optimal strategy is to find the $\theta_{m}$ closest to $\theta$
and to send in the program state $|\theta_{m}\rangle$ corresponding to that value.  If we do so we
find that 
\begin{equation}
F_{proc}(U(\theta ),U(\theta_{m}))> \cos^{2}\left(\frac{\pi}{N}\right) \sim 1- \left(\frac{\pi}{N}\right)^{2} .
\end{equation}
Rather than determining which $\theta_{m}$ is the best one to use, a simpler procedure is just to
use the program state $|\theta\rangle$.  There should be some cost to doing this, and, indeed, we 
find that the process fidelity in this case is approximately $1-(2/N)$.  The optimal program has an
error that goes like $1/N^{2}$ while the simpler procedure gives an error of $1/N$.  Determining 
whether the extra accuracy is worth the extra work in determining the best program would depend
on the application.

Now let us return to probabilistic programmable devices.  Suppose our data system is a qubit, and we
want to implement the one parameter group we mentioned earlier, $U(\alpha )=
\exp (i\alpha \sigma_{z})$, where $0\leq \alpha < 2\pi$.  This can be accomplished with a success
probability of $1/2$ by using a qubit program and a controlled-NOT gate.  
As was noted before, the controlled-NOT gate
has two inputs, a control input and a target input.  The state of the control qubit is not changed, 
and if the state of the control qubit is $|0\rangle$, neither is the state of the target qubit.  However,
if the control qubit is in the state $|1\rangle$, then the operator $\sigma_{x}$ is applied to the 
target qubit.  It is, in fact, a controlled-U gate with the two unitary operators being the identity and
$\sigma_{x}$.  In our case, the target qubit is the program and the control qubit is the data.  The
program states are
\begin{equation}
|\Xi (\alpha )\rangle = \frac{1}{\sqrt{2}}(e^{i\alpha} |0\rangle + e^{-i\alpha}|1\rangle ) .
\end{equation}
If the data state input is $|\psi\rangle$, the output of this processor is then
\begin{equation}
|\Psi_{out}\rangle = \frac{1}{\sqrt{2}}(U(\alpha )|\psi\rangle |0\rangle + U^{-1}(\alpha )|\psi\rangle
|1\rangle ) .
\end{equation}
By measuring the program state output in the basis $\{ |0\rangle , |1\rangle \}$,and keeping the 
result only if we get $|0\rangle$, which happens with a probability of $1/2$, we obtain the data
state output $U(\alpha )|\psi\rangle$, which is the desired result.

A closely related programmable device has been recently realized experimentally \cite{micuda}.
It carries out the transformation
\begin{equation}
\alpha |0\rangle + \beta |1\rangle \rightarrow \alpha |0\rangle + e^{i\phi} \beta |1\rangle ,
\end{equation}
where the angle $\phi$ is encoded in a second qubit.
The qubits are polarization states of photons, with $|H\rangle$ representing a horizontally
polarized photon and $|V\rangle$ representing a vertically polarized one.  A polarizing beam
splitter, which transmits horizontally polarized photons and reflects vertically polarized ones is
the main component of the device.  The beam splitter has two input modes, which we shall 
label $1$ and $2$, and two output modes, which we shall also denote as $1$ and $2$.  For a 
photon incident in input mode $1$ we would have $|H\rangle_{1}\rightarrow |H\rangle_{1}$ and
$|V\rangle_{1}\rightarrow |V\rangle_{2}$.  Input mode $2$ behaves similarly.  If two photons, one 
in the state $\alpha |H\rangle_{1} + \beta |V\rangle_{1}$ (data) and the other in the state 
$(1/\sqrt{2})(|H\rangle_{2} + e^{i\phi}|V\rangle_{2} )$ (program) are incident on the polarizing beam 
splitter, then in the cases in which a single photon emerges from each output, which happens with
a probability of $1/2$, the conditional output state is
\begin{equation}
|\psi_{out}\rangle = \frac{1}{\sqrt{2}}( \alpha |H\rangle_{1}|H\rangle_{2} + e^{i\phi}\beta  
|V\rangle_{1} |V\rangle_{2} ) .
\end{equation}
If we measure the second photon in the $|\pm\rangle = (1/\sqrt{2})(|H\rangle \pm |V\rangle )$ basis,
then the remaining photon is in either the state $(1/\sqrt{2})(|H\rangle +e^{i\phi} |V\rangle )$, if our
measurement result was $|+\rangle$, and $(1/\sqrt{2})(|H\rangle - e^{i\phi} |V\rangle )$, if our
measurement result was $|-\rangle$.  If we obtain the result $|-\rangle$ we can apply a correcting
operation on the remaining qubit that sends $|H\rangle \rightarrow |H\rangle$ and $|V\rangle
\rightarrow -|V\rangle$.  The final result is that this device implements the transformation
\begin{equation}
\alpha |H\rangle + \beta |V\rangle \rightarrow \alpha |H\rangle + e^{i\phi} \beta |V\rangle ,
\end{equation}
with a probability of $1/2$.

Suppose that we want to increase the probability of a successful outcome.  One possibility is to
try again if get the wrong result of our measurement on the program state \cite{preskill,vidal}.
If we obtained the result $|1\rangle$ from our measurement, then the data qubit is in the state
$U^{-1}(\alpha )|\psi\rangle$.  We can take this quibit and run it through the processor again,
but this time use the program $|\Xi (2\alpha )\rangle$.  If we do so, the output state is
\begin{equation}
|\Psi_{out}^{\prime}\rangle = \frac{1}{\sqrt{2}}(U(\alpha )|\psi\rangle |0\rangle 
+ U^{-1}(3\alpha )|\psi\rangle |1\rangle ) .
\end{equation}
We again measure the program state and keep the result if we get $|0\rangle$.  This again happens
with a probability of $1/2$.  Adding this second step has increased our overall success probability
to $3/4$, and the procedure can be repeated to bring the success probability as close to one as
we wish.  What we need to do this, however, is a collection of qubits in the proper program states,
that is, besides a qubit in the state $|\Xi (\alpha)\rangle$, we need an additional one in the 
state $|\Xi (2\alpha)\rangle$.

We can also accomplish the same thing by enlarging our program space \cite{vidal}.  Our data
space still consists of one qubit, but the program space now contains two qubits.  Let us label the
three inputs, input $1$ being the data input, input $2$ the first program input and input $3$ the second
program input.  The processor now consists of two gates.  The first is a controlled-NOT gate
whose control qubit is qubit $1$ and whose target qubit is qubit $2$.  The second gate is a Toffoli 
gate.  This gate has two control qubits and one target qubit.  The states of the control qubits are not
changed, and if they are in the states $|0\rangle |0\rangle$, $|0\rangle |1\rangle$, or 
$|1\rangle |0\rangle$, neither is the state of the target qubit.  However, if they are in the state
$|1\rangle |1\rangle$, then $\sigma_{x}$ is applied to the target qubit.  In our processor, qubits
$1$ and $2$ are the control qubits and qubit $3$ is the target qubit.  The input state is
$|\psi\rangle_{1}|\Xi (\alpha )\rangle_{2} |\Xi (2\alpha )\rangle_{3}$, and the output state is
\begin{equation}
|\Psi_{out}^{\prime\prime}\rangle = \frac{1}{2}[U(\alpha )|\psi\rangle_{1} (|0\rangle_{2}|0\rangle_{3} 
+|0\rangle_{2}|1\rangle_{3}+|1\rangle_{2}|0\rangle_{3})+ U^{-1}(3\alpha )|\psi\rangle |1\rangle_{2}
|1\rangle_{3} ] .
\end{equation}
At the output we measure the program qubits in the computational basis and keep the data state
output if we get $|0\rangle |0\rangle$, $|0\rangle |1\rangle$, or $|1\rangle |0\rangle$.  If we do,
the data output is in the state $U(\alpha )|\psi\rangle$, and we have achieved our goal.  This 
happens with a probability of $3/4$.  By increasing the dimension of the program space further,
we can increase our probability of success.  We have, therefore, two strategies for increasing the
success probability for a probabilistic processor.

\section{Why quantum programs?}
The programs in the quantum processors we have been discussing have been quantum states.  One
might wonder whether this is necessary and whether classical programs would suffice.  That is,
one could have gates that can perform a number of operations, but the selection of which operation
they do perform is governed by a classical input.  Do quantum programs provide an advantage?
There are several scenarios that suggest themselves for which quantum programs would be
useful.  One is that the information on which the program is based is intrinsically quantum.  We
shall explore an example of this situation when we discuss programmable state discriminators.
This could also occur if the program is the result of an earlier quantum computation.  A second
situation is one in which we would like to apply quantum information processing techniques, such 
as a Grover search, to programs.  In that case, the programs must be quantum.

Let us first consider programmable state discriminators.  The first such device was proposed by
Bu\v{z}ek and Du\v{s}ek \cite{buzek3}.  Here we will discuss a different version, which is a type
of universal state discriminator \cite{bergou2}.  So far, when discussing state discriminators, we have 
assumed we knew the set of states we were trying to discriminate among.  This knowledge was 
built into the discriminator.  The resulting discriminator is useful for discriminating states from that
particular set, but it is not useful for discriminating among members of other sets of states.  Suppose,
however, that we would like a discriminator that would work for any set of states, i.e. a universal
discriminator.  In that case, we have to provide information about the set of possible states as well 
as the quantum system whose state we want to determine with the machine.  The information about
the set of possible states will be the program.

Let us consider the simplest version of such a device.  It will unambiguously discriminate between
two different qubit states.  The program consists of two states, one in each of the states we want
to discriminate between, which we shall call $|\psi_{1}\rangle$ and $|\psi_{2}\rangle$.  The data
qubit is in either $|\psi_{1}\rangle$ or $|\psi_{2}\rangle$, and we would like to know which.  What
the machine does is implement a POVM, which takes advantage of the symmetry of the three-qubit
input state.  Let us call the program inputs $a$ and $b$, and the data input $c$.  Our task is to
discriminate between the states
\begin{eqnarray}
|\Psi_{1}\rangle & = & |\psi_{1}\rangle_{a} |\psi_{2}\rangle_{b} |\psi_{1}\rangle_{c}  \nonumber \\
|\Psi_{2}\rangle & = & |\psi_{1}\rangle_{a} |\psi_{2}\rangle_{b} |\psi_{2}\rangle_{c}  .
\end{eqnarray}
Note that in $|\Psi_{1}\rangle$ the first and third qubits are in the same state, while in 
$|\Psi_{2}\rangle$ the second and third qubits are in the same state.  Therefore, if we project
the three-qubit input state onto the antisymmetric subspace of qubits $a$ and $c$, and we
get a nonzero result, then we know that qubit $c$ was in the state $|\psi_{2}\rangle$.  Similarly, 
if we project qubits $b$ and $c$ onto the antisymmetric subspace of two qubits, and we get
a nonzero result, then we know that qubit $c$ was in the state $|\psi_{1}\rangle$.  There will
also be a ``don't know'' result in which the measurement fails, and we want to minimize the
probability of obtaining this result.  If the two states are equally likely, and averaging over 
$|\psi_{1}\rangle$ and $|\psi_{2}\rangle$, since we do not know what they are, we find
that the optimal probability of identifying the input data state is $1/6$.  Note that in this case,
the information contained in the program was quantum information, in particular, it consisted 
of examples of quantum states, and this necessitated the program itself being quantum.

Now let us look at an example in which it is useful to apply quantum information processing
techniques to quantum programs.  In order to do so, we first need to explain the quantum
search algorithm due to Lov Grover \cite{grover}.  We have a black box that evaluates a 
Boolean function.  A Boolean function is one whose value is either zero or one.
We send in an input, which is an $n$-digit binary number, $x$, and the output of the box is 
$f(x)$.  This particular function is zero on all inputs except one, which we shall call $x_{0}$,
and $f(x_{0})=1$.  Our object is to find $x_{0}$ with a minimum number of uses of the black box.

Classically, we simply send in different inputs to the black box until we find one that gives one as
an output.  On average we will have to make $2^{(n-1)}$ tries.  The Grover algorithm works in a
completely different way, and its result is a considerable improvement over the classical one.  It
starts with an input state that is an equal superposition of all possible input values.  By successively
applying the black box followed by an operator Grover called ``inversion about the mean'' 
approximately $2^{n/2}$ times, the initial state is rotated into the state $|x_{0}\rangle$, and then
one simply measures this state in the computational basis to find out what $x_{0}$ is.  Note that
the black box was only used $2^{n/2}$ times in this case, which means that the number of evaluations
in the Grover algorithm is approximately the square root of the number of evaluations that are
necessary in the classical case.

Now consider the following problem \cite{bonanome}.  We have a set of $M$ permutations on 
$N$ objects.  In particular let $X=\{ k | k=0,1,\ldots N-1\}$ be the set of objects being permuted, 
and let $S  = \{ \sigma_{j} | j=1,2,\ldots M\}$ be the set of permutations.  For some specified 
$k_{0},k_{1}\in X$, we are promised that there is one $\sigma_{j}\in S$ such that $\sigma (k_{0})
=k_{1}$, and we want to find which permutation satisfies this property.  A variant of this problem,
determining whether there is a  $\sigma_{j}\in S$ such that $\sigma (k_{0}) =k_{1}$, can be used
to attack the conjugacy problem in group theory.  If $G$ is a group, and $g_{1},g_{2}\in G$, we 
would like to know whether $g_{1}$ and $g_{2}$ are conjugate to each other, that is, whether
there is an $h\in G$, such that $g_{2}=hg_{1}h^{-1}$.  The connection between this problem
and the one involving the permutations is provided by realizing that the automorphism $\alpha_{h} :
G\rightarrow G$ given by $\alpha_{h} (g) = hgh^{-1}$ is just a permutation on $G$.  Thus,
the conjugacy problem is reduced to determining whether there is an $\alpha_{h}$ such that
$\alpha_{h}(g_{1})= g_{2}$.

We suppose we have a quantum processor, which acts on the Hilbert space $\mathcal{H}_{X}
\otimes \mathcal{H}_{S}$, where $\mathcal{H}_{X}$ is spanned by the orthonormal basis 
$\{ |k\rangle_{X} | k=0,1,\ldots N-1\}$ and $\mathcal{H}_{S}$ is spanned by the orthornormal
basis $\{ |j\rangle_{S}  | j=1,2,\ldots M\}$.  We regard $\mathcal{H}_{S}$ as the program space, 
and $\mathcal{H}_{X}$ as the data space.  The processor acts as follows
\begin{equation}
|j\rangle_{S} |k\rangle_{X} \rightarrow |j\rangle_{S} U_{j}|k\rangle_{X}  ,
\end{equation}
where $U_{j}|k\rangle_{X}=|\sigma_{j}(k)\rangle_{X}$.  Once we have this processor, we can
do a Grover search on the programs in order to find the permutation that satisfies 
$\sigma (k_{0}) =k_{1}$.  This will require approximately $\sqrt{M}$ uses of the processor, whereas
classically $M$ uses would be required.  It is the fact that the programs are quantum states that
allows us to search among them by using a quantum search procedure.

\section{Conclusion}
As we have seen, quantum machines have been developed for a number of information processing
tasks.  Cloners move quantum information around and discriminators allow one to distinguish among
nonorthogonal quantum states.  Discriminators can be generalized to distinguish between 
nonorthogonal subspaces as well \cite{bergou3}.  In addition, we have seen that it is possible to
construct programmable quantum machines, which are capable of performing a number of different
tasks. 

The capabilities of programmable machines are still not well understood.  We concentrated
mainly on processors that implement unitary operators, but, as we saw processors can also
implement more general maps.  Some families of maps, for example, those in Eq.\ (\ref{procmap}),
can be implemented with a finite-dimensional program space, while others, such as a one-parameter
unitary group, cannot.  What determines whether a set of maps can be programmed with a finite-
dimensional program?  Another issue is the equivalence of programmable processors.  Suppose
we have two processors, both of which can perform the same set of operations but they do so with
different programs.  This could happen, for example, if the one of the processors differed from the
other simply by having a fixed unitary gate at the input to its program register.  Given two processors,
is there a simple way of telling whether or not the set of operations they can implement is the same?
These are only two questions about the properties of quantum processors, and we suspect there are
many more.  

\section*{Acknowledgments}
This work was supported by the European Union projects HIP and QAP, by Slovak grant agencies
APVV and VEGA via projects RPEU-0014-06 and 2/0092/09, respectively.

\end{document}